\documentclass[12pt]{article}
\usepackage{amsmath,amsfonts, amssymb,graphicx,geometry,authblk,color,soul,comment,hyperref} 
\usepackage{bm}

\begin{document}
\title{\textbf{Adhesion of a nematic elastomer cylinder}}

\author{Ameneh Maghsoodi$^{1}$}
\author{ Kaushik Bhattacharya$^{2}$}
\affil{$^{1}$ Department of Aerospace and Mechanical Engineering, University of Southern California, Los Angeles, CA 90089, USA}
\affil{$^{2}$Division of Engineering and Applied Science, California Institute of Technology, Pasadena, CA 91125, USA}
\maketitle

\begin{abstract}

\noindent Liquid crystal elastomers are cross-linked elastomer networks with liquid crystal mesogens incorporated into the main or side chain. Polydomain liquid crystalline (nematic) elastomers exhibit unusual mechanical properties like soft elasticity, where the material deforms at nearly constant stress, due to the reorientation of mesogens. In this paper, we use numerical simulation to study the implication of the remarkable elastic softness on a classical problem of adhesion. This study reveals that the soft elasticity of nematic elastomers dramatically affects the interfacial stress distribution at the interface of a nematic elastomer cylinder adhered to a rigid substrate. The stress near the edge of nematic cylinder under tensile load deviates from the singular behavior predicted for linear elastic materials, and the maximum normal stress reduces dramatically.  Moreover, the location of maximum interfacial stress shifts from the edge to the center of the nematic cylinder when the applied tensile force goes beyond a critical value. We discuss the implications for adhesion. The results are consistent with the available experimental data.
 
\end{abstract}
\maketitle

\begin{figure}
\begin{center}
  \includegraphics[width=6.5in]{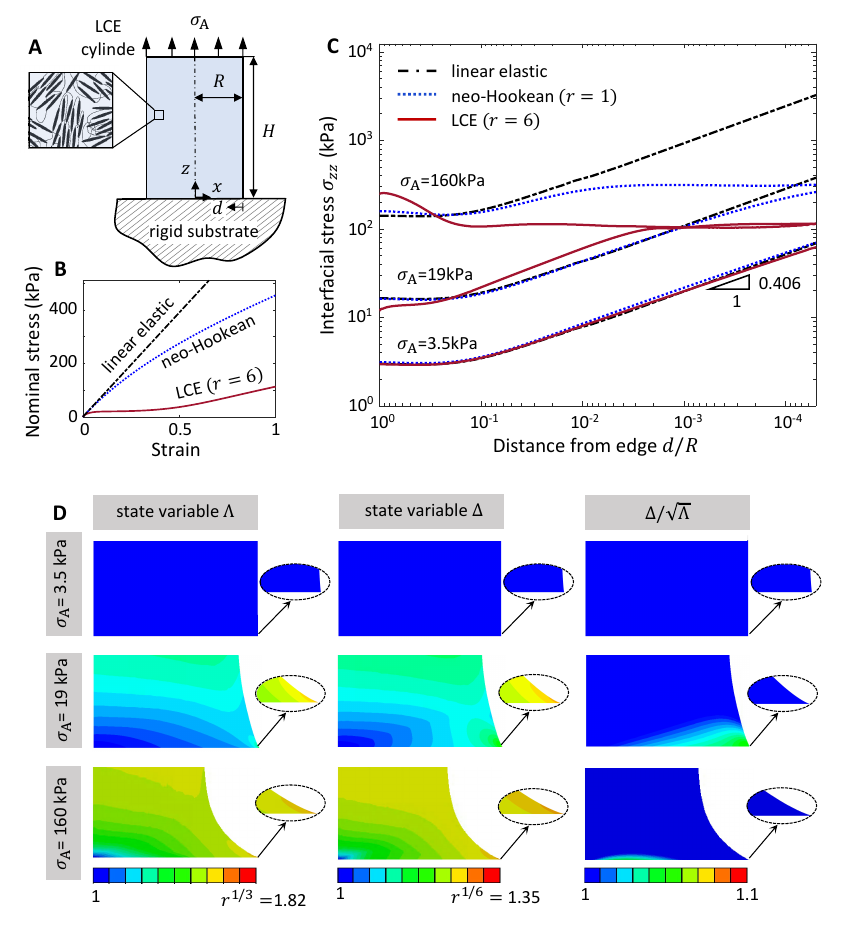}
  \caption{(A) Schematic of an isotropic-genesis polydomain nematic LCE cylinder adhered to a rigid surface. (B) The uniaxial nominal stress-strain curves for linear elastic material, neo-Hookean hyperelastic ($r=1$) material, and LCE ($r=6$). (C) Distribution of interfacial normal stress $\sigma_{zz}$ along the adhered interface of the cylinder in logarithmic scale for materials in (B). (D) Distribution of the state variables $\Lambda$, $\Delta$, and ratio $\Delta/\sqrt{\Lambda}$ for LCE with $r=6$ near the adhered region at different applied loads $\sigma_A$. The insets show the singular region at the edge of the cylinder.} 
\label{fig:fig1}
\end{center}
\end{figure}

\noindent \textbf{Introduction} \\
Liquid crystal elastomers (LCEs) are cross-linked elastomer networks with liquid crystal molecules, also known as mesogens,  incorporated into the underlying polymer chains. Mesogens are stiff, rod-like molecules that respond to temperature by changing their orientation distribution. At high temperatures $T>T_\mathrm{ni}$  ($T_\mathrm{ni}$ denotes the nematic to isotropic transition temperature), the LCE is in the isotropic state where the mesogens are randomly oriented. At lower temperatures $T<T_\mathrm{ni}$, the LCE is in the nematic state where the mesogens are aligned along a preferred direction. The degree of order observed in the mesogens determines the degree of anisotropy. When an isotropic-genesis LCE (one that is cross-linked in the isotropic state) is cooled down, it undergoes a phase transition from its isotropic state to a nematic state and forms an isotropic-genesis polydomain nematic LCE with domains on the order of 1-2 $\mu$m \cite{clarke1998texture}; see Fig. \ref{fig:fig1}A. 

A fascinating characteristic of isotropic-genesis nematic LCEs is the soft elasticity behavior: when subjected to an external uniaxial tension, the material stretches at almost zero stress, resulting in a soft plateau region in the stress-strain curve \cite{biggins2009supersoft,biggins2012elasticity,urayama2009polydomain}. This phenomenon is attributed to the reorientation of the mesogens through polydomain-monodomain transition \cite{clarke1998texture,biggins2009supersoft,biggins2012elasticity,fridrikh1999polydomain}. Practically, nematic elastomers exhibit a non-ideal `semi-softness' response due to the presence of internal constraints, leading to an initial linear elastic regime before the stress plateau. This produces a non-zero stress plateau until the full chain re-alignment is achieved. Several microscopic mechanisms contribute to this non-ideal semi-softness response including the polydispersity of network chains \cite{verwey1997compositional}, the effect of anisotropic cross-linkers \cite{verwey1997nematic}, and the entanglement of nematic chains \cite{kutter2001tube}. Recent research indicates that the soft elasticity of LCE makes its mechanical behavior differ dramatically from that of rubber in various problems including the wrinkling of thin sheets \cite{plucinsky2017microstructure}, energy absorption in impact \cite{jeon2022synergistic}, and Hertz contact \cite{MAGHSOODI2023102060}. Interestingly, recent experiments \cite{farre2022dynamic,ohzono2019enhanced} exhibit that the adhesion force between glass and a polydomain nematic LCE is higher than that between glass and silicone rubber. In this paper, we study how the soft elasticity of nematic LCE contributes to a stronger adhesion. 

Consider a flat-ended cylinder perfectly attached to a rigid substrate at one end; see Fig.~\ref{fig:fig1}A.    If the cylinder is linear elastic and subjected to a tensile load at the other end, the maximum normal stress on the cylinder-substrate interface occurs at the edge where the cylinder touches the substrate. In fact, the normal stress distribution is singular at this edge and is of the form $\sigma=Kd^n$, where the intensity $K$ depends on the applied load, $d$ is the distance from the edge, and $n= -0.406$ \cite{khaderi2015detachment}.  This stress singularity results in a crack being initiated at the edge, and this eventually leads to the failure of adhesion.  In this work, we examine the stress distribution on an LCE flat-ended cylinder perfectly attached to a rigid substrate at one end and subject to an applied tensile load at the other. We find that the soft elasticity dramatically changes the interfacial stress distribution in the LCE cylinder.  The stress is no longer singular at the edge and the location of the maximum stress shifts to the interior.  We discuss the implications for adhesion and compare the results with available experimental data.\\

\vspace{0.15cm}
 \noindent \textbf{Model}  \\
 We consider a cylinder under normal loading as shown in Fig. \ref{fig:fig1}A.   The bottom surface of the cylinder is fixed in all directions to model adhesion to a rigid flat substrate, while the top interface of the cylinder is displaced uniformly in the axial or z-direction with no displacement allowed in other directions modeling the fact that the cylinder is bonded to a stiff support plate at the top.  The lateral surfaces are traction-free.  We consider a relatively long cylinder to eliminate end effects on the stress distribution at the interface.
 
 We model the cylinder as a 2D axisymmetric model in the commercial finite element package ABAQUS Standard \cite{abaqus}.  The cylinder is discretized using four-node bilinear axisymmetric hybrid elements.   The mesh near the contact interface and the free edge is refined, and mesh convergence is verified with further refinement.  We verify that we resolve the singularity at the edge by plotting the stress on a semi-log plot and verifying the slope against known theoretical values.  Further, we use the same mesh for all our simulations.
 
 We use the constitutive model for an isotropic-genesis polydomain nematic elastomer developed by Lee {\it et al.}\cite{LEE2023105369}. We provide a brief overview of the model here.   This model introduces two scalar state variables $\Lambda$ and $\Delta$ that describe the spontaneous deformation associated with the local domain pattern. These are closely related to local polydomain order parameters: $\Lambda$ with the degree of orientation $S$, and $\Delta$ with $S+X$ where $X$ is the degree of biaxial orientation. These state variables describe the spontaneous change in material metric (the Cauchy-Green stretch due to domains) $\bm{G} = \bm{P} \ \text{diag}(\Lambda^2,\Delta^2/\Lambda^2,1/\Delta^2)\bm{P}^T$ where $\bm{P}$ is a rotation matrix, and $\Lambda$ and $\Delta$ can take values in the region $\{( \Delta \le r^{1/6}, \Delta \le \Lambda^2, \Delta \ge \sqrt{\Lambda}\}$ where $r$ is the chain anisotropy parameter (related to the degree of nematic order $Q$).  A monodomain has $\Lambda=r^{1/3}$ and $\Delta=r^{1/6}$ so that $\bm{G}$ is the step-length tensor $\bm \ell$ of the neo-classical theory \cite{WTbook}, and an isotropic polydomain state where the nematic directors are equidistributed has $\Lambda=\Delta=1$ so that $\bm{G}$ is identity.  The biaxial polydomain state where all the nematic directors are confined to a plane but equidistributed in the plane has $\Lambda=r^{1/12}$ and $\Delta=r^{1/6}$ so that $\bm{G}= \bm{P} \ \text{diag} (r^{1/12}, r^{1/12}, r^{-1/6})\bm{P}^T$. The model postulates a coarse-grained free energy $W=W_e+W_r$ where $W_e = \frac{1}{2}\mu [\text{tr} (\bm{F}^T \bm{G}^{-1} \bm{F}) - 3]$ is the entropic energy in the polymer chains for a deformation gradient $\bm{F}$ relative to an isotropic reference state, with $\mu$ the rubber modulus, and $W_r = C(\Delta-1)/(r^{1/6}-\Delta )^k$ is the energy of domain patterns required to overcome fluctuations. The deformation is determined by the equation of mechanical equilibrium while the state variables evolve according to overdamped dynamics $\alpha_\Lambda \dot \Lambda = -\partial W/\partial \Lambda, \ \alpha_\Delta \dot \Delta = - \partial W/\partial \Delta$. The model has been validated against experiments and verifiably implemented as a UMAT in the finite element package ABAQUS \cite{LEE2023105369}. The typical material properties we use in our simulations are $\mu=0.26$MPa, $C=0.6$kPa, $\alpha_\Delta=30$MPa.s, $\alpha_\Lambda=0.01\alpha_\Delta$, $k=2$, and $r=6$ for a nematic LCE cylinder. Note that we can include the neo-Hookean rubber into this model by setting $r=1$. \\

\noindent{\noindent \textbf{Results and Discussion} \\
\noindent\textbf{Interfacial stress distribution.} Fig.~\ref{fig:fig1}C displays the distribution of normal stress along the adhered interface for a linear elastic material, a neo-Hookean hyperelastic material (LCE with $r$=1), and a nematic LCE with $r=6$ at various values of the applied load. We present the results in a logarithmic scale to highlight the details of the stress singularity near the edge of the cylinder. For the linear elastic cylinder, a significant stress concentration is generated at the edge of the interface in the form of  $\sigma_{zz}=Kd^{n}$ with $n=-0.406$ as anticipated in the classical theory \cite{khaderi2015detachment}, thereby providing a verification of the numerical method.  This singularity exists at all values of the applied load, with the intensity $K$ proportional to the applied load.  A neo-Hookean hyperelastic cylinder ($r=1$) follows the linear elastic theory for small values of the applied load (e.g., 3.5 kPa), but then deviates from it at larger values of the applied load.  At $\sigma_A=160$kPa, there is no singularity at the edge.  Further, the value of the normal stress at the edge is significantly less than that in a linear elastic material, but looks similar away from the edge except it is slightly elevated at the center (to give the same average stress).

Finally, we turn to the LCE cylinder with $r=6$. At low applied loads ($\sigma_A=3.5$kPa), the stress distribution follows the linear elastic theory with an exponent $n=-0.406$. This is because of the initial elastic regime in the stress-strain response of the non-ideal LCE; see the stress-strain curve in Fig.~\ref{fig:fig1}B.  However, it soon deviates as we increase the load.  The singularity at the edge vanishes and the level of stress at the edge is significantly reduced compared to the other two materials, at $\sigma_A=19$kPa.  As the load increases further, the stress distribution is still regular.  Further, the value at the edge remains unchanged despite the increased applied load and increases in the center of the cylinder instead, see $\sigma_A=160$kPa.  Thus the stress at the edge is significantly smaller in the LCE compared to that in the other two cases, but higher in the center. 

To gain insight into the reason for this dramatically different stress distribution in the LCE, we study the domain pattern and its evolution.  Fig.~\ref{fig:fig1}D shows the distribution of the state variables $\Lambda$ and $\Delta$, and the ratio $\Delta/\sqrt{\Lambda}$ in the vicinity of the adhered region at three different applied loads $\sigma_A$.   The color scale in Fig.~\ref{fig:fig1}D for $\Lambda$ and $\Delta$ are chosen so that blue corresponds to the smallest value ($1$ for both) while red corresponds to the largest theoretical value ($r^{1/3} = 1.82$ for $\Lambda$, $r^{1/6} = 1.35$ for $\Delta$).   However, the color scale for  $\Delta/\sqrt{\Lambda}$ is chosen to be limited to be close to 1 (the possible maximum value for $\Delta/\sqrt{\Lambda}$ is $r^{1/8} = 1.25$ but our scale only goes to 1.1).   As the applied load $\sigma_A$ increases, $\Lambda$ evolves significantly, especially near the edge of the cylinder with the maximum value at the edge. This maximum $\Lambda$ at the edge reaches the saturation value of $\approx 1.71$ at the higher applied load ($\sigma_A=160$kPa) and this is close to the theoretical maximum value of 1.82 (the material hardens significantly as it approaches the maximum value in the constitutive model). We observe that $\Delta$ also evolves and reaches the value of $\approx 1.31$ (close to the maximum values of $1.35$) at the edge of the cylinder at the higher load. However, the ratio $\Delta/\sqrt{\Lambda}\approx 1$ everywhere along the adhered interface in all cases. The ratio $\Delta/\sqrt{\Lambda} = 1$ indicates a pure uniaxial deformation.   Therefore, we conclude that the domain pattern evolves to maintain an uniaxial state of deformation along the adhered interface. Further, $\Lambda \approx r^{1/3}$ and $\Delta \approx r^{1/6}$  at the edge, and thus the LCE is almost in a monodomain state.   In other words, the polydomain-monodomain transition suppresses the stress singularity at the edge in an LCE cylinder. 

Figure \ref{fig:fig2}A shows the maximum interfacial normal stress $\sigma_{zz}^{max}$ and the corresponding radial position where it is attained for different applied loads $\sigma_A$. The maximum interfacial stress for the neo-Hookean material remains at the edge of the cylinder (a$_1$-e$_1$), as exemplified in Fig.~\ref{fig:fig1}C. However, LCE exhibits a significantly different trend.  At low applied load, the maximum interfacial stress is located at the edge (a$_6$-c$_6$), but the location of maximum stress shifts from the edge to the center of the cylinder (d$_6$-e$_6$) above a transition applied load $\sigma_A^t$; see Fig.~\ref{fig:fig1}C. As shown in Fig.~\ref{fig:fig2}B, the interfacial stress at the transition applied load $\sigma_{A}^{t}$ is almost uniform along the adhered interface.   \\

\begin{figure}
\begin{center}
  \includegraphics[width=6.5in]{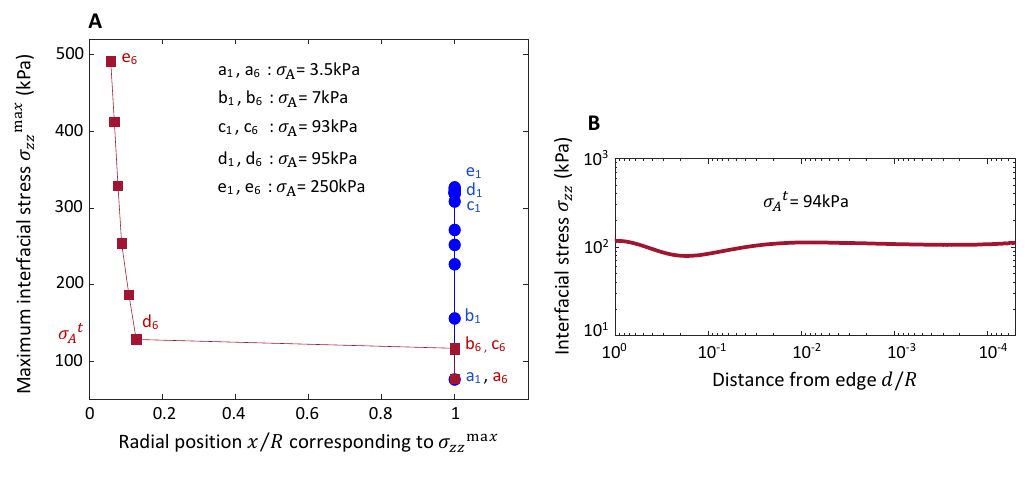}
  \caption{(A) Maximum interfacial stress $\sigma_{zz}^{max}$ vs. its corresponding radial position along the adhered interface of the cylinder at different applied loads $\sigma_A$. The red square data correspond to the nematic LCE cylinder with $r=6$ and the blue circle data correspond to the isotropic cylinder ($r=1$). (B) Distribution of interfacial normal stress $\sigma_{zz}$ along the adhered interface of the cylinder at transition applied load $\sigma_A^t=94$kPa for LCE with $r=6$.  } 
\label{fig:fig2}
\end{center}
\end{figure}

\noindent\textbf{Adhesion.}  The failure adhesion in a cylinder of sufficiently large diameter occurs by a process of nucleation of a crack either at a pre-existing flaw or stress singularity followed by growth.  Tann\'e {\it et al.} \cite{tanne} (also \cite{hsueh}), based on an extensive study of experimental and computational observations, proposed a unified criterion for stress nucleation at a point where the opening stress is locally of the form $\sigma \sim K d^n $ (so $n=0$ for a non-singular stress field, $n=-0.406$ at the edge of an adhered cylinder, and $n=-0.5$ for a pre-existing crack).  A crack nucleates when
\begin{equation}
K = K_c:= K_{Ic}^{-2n} \sigma_c^{1+2n}
\end{equation}
where $K_{Ic}$ is the fracture toughness and $\sigma_c$ is the maximum tensile strength. Note that $K=K_{Ic}$ in the presence of a pre-existing crack and $\sigma=\sigma_c$ in the absence of a singularity in agreement with classical fracture mechanics.

We now apply this criterion to the current problem of cylinder adhesion.  In a linear elastic cylinder, the stress is singular at the corner with $n=-0.406$, and therefore, failure initiates at the corner. Therefore, the stress intensity factor $K$ determines the adhesion strength of the cylinder \cite{balijepalli2016numerical,balijepalli2017numerical,luo2020adhesion}; the lower stress intensity at the edge of the cylinder leads to a higher adhesion strength.  In a neo-Hookean cylinder, the stress is initially singular at the edge, but then becomes regular.  At that point, though the highest stress occurs at the edge (Fig. \ref{fig:fig2}), the stress is quite uniform (Fig. \ref{fig:fig1}).  It is also known that in shorter cylinders, the interior stress can also increase \cite{gent}.  For these reasons, failure may occur at the edge or in the interior depending on the specific dimensions and properties \cite{yao2006mechanics}.  However, the failure will occur at significantly higher values of the applied load compared to a linear elastic material of similar properties.

In an LCE cylinder, the stress is singular at the edge for small applied loads, but the stress intensity is insufficient to cause failure.  The singularity decreases and eventually goes away.  Further, the levels of stress are significantly lower in an LCE cylinder compared to that of the neo-Hookean cylinder (Figs. \ref{fig:fig1} and \ref{fig:fig2}).  In other words, there are two mechanisms for the suppression of failure -- lack of a stress singularity at the edge, and significantly reduced levels of stress.  This leads to a significant increase in the adhesive strength of an LCE cylinder compared to that of a neo-Hookean one. This is consistent with the experimental observation by Farre-Kaga {\it et al.} \cite{farre2022dynamic} where they tested the adhesion of the polydomain nematic LCE via the probe-tack experiment (rigid cylinder and LCE substrate).    Further, since the location of the highest stress is in the interior, we expect the failure to initiate in the interior even for long cylinders.\\

\begin{figure}
\begin{center}
 \includegraphics[width=6.5in]{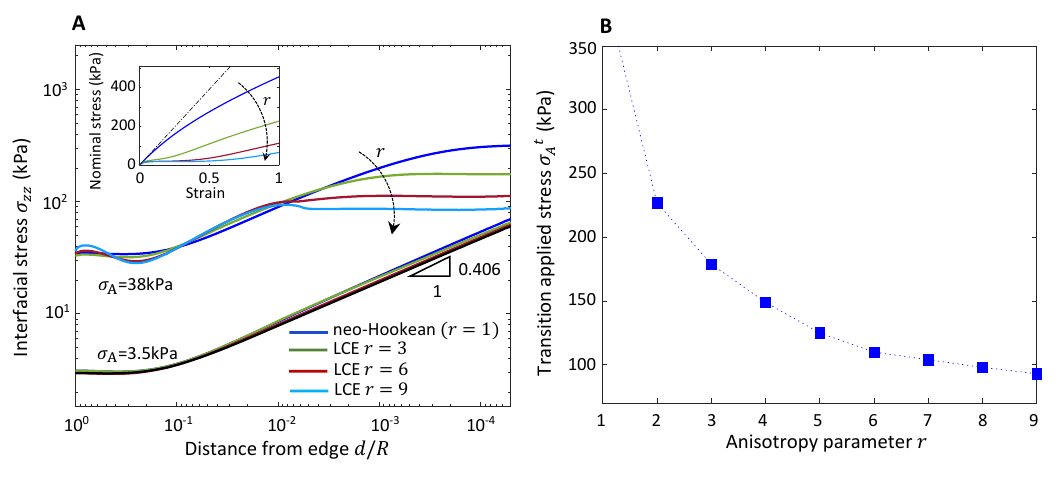}
  \caption{(A) Distribution of interfacial normal stress $\sigma_{zz}$ along the adhered interface of the cylinder in logarithmic scale at different anisotropy $r$. The inset presents the corresponding uniaxial nominal stress-strain curves. (B) Transition applied load $\sigma_{A}^t$ as a function of anisotropy parameter $r$.} 
\label{fig:fig3}
\end{center}
\end{figure}

\noindent\textbf{Effect of temperature.} Temperature $T$ significantly affect anisotropy parameter $r$. At $T_{ni}<T$, the LCE is in the isotropic state with $r=1$, and at $T<T_{ni}$, the LCE is nematic with $r>1$. Fig.~\ref{fig:fig3}A compares the interfacial stress distribution at different anisotropy parameters $r$ (equivalently, different temperatures) with the corresponding uniaxial stress-strain curves in Fig.~\ref{fig:fig3}(inset). At low load ($\sigma_A=3.5$kPa), all cases follow the linear elastic theory with the same stress distribution along the interface including a singularity at the edge with the exponent $n=-0.406$.   The singularity vanishes for all the studied cases at the higher applied load ($\sigma_A=38$kPa).  However, the value of the stress at the edge decreases with increasing $r$ (decreasing temperature).   These lead us to conclude that lower temperature would lead to stronger adhesion, and this is consistent with experimental observations of Ohzono {\it et al.}  \cite{ohzono2019enhanced}.  Fig.~\ref{fig:fig3}B shows the transition applied load $\sigma_A^t$ at which the point of maximum stress shifts from the edge to the center.  We observe that this transition occurs at a smaller load for larger $r$ (lower temperature).  Therefore, we anticipate an interior failure at lower temperatures.
\subsection*{Acknowledgements}
This work has been funded by the US Office of Naval Research (MURI grant N00014-18-1-2624)

\bibliographystyle{unsrt}
\bibliography{references}

\begin{thebibliography}{10}

\bibitem{clarke1998texture}
S.M. Clarke, E.M. Terentjev, I.~Kundler, and H.~Finkelmann.
\newblock Texture evolution during the polydomain-monodomain transition in nematic elastomers.
\newblock {\em Macromolecules}, 31(15):4862--4872, 1998.

\bibitem{biggins2009supersoft}
J.S. Biggins, M.~Warner, and K.~Bhattacharya.
\newblock Supersoft elasticity in polydomain nematic elastomers.
\newblock {\em Physical review letters}, 103(3):037802, 2009.

\bibitem{biggins2012elasticity}
J.S. Biggins, M.~Warner, and K.~Bhattacharya.
\newblock Elasticity of polydomain liquid crystal elastomers.
\newblock {\em Journal of the Mechanics and Physics of Solids}, 60(4):573--590, 2012.

\bibitem{urayama2009polydomain}
K.~Urayama, E.~Kohmon, M.~Kojima, and T.~Takigawa.
\newblock Polydomain- monodomain transition of randomly disordered nematic elastomers with different cross-linking histories.
\newblock {\em Macromolecules}, 42(12):4084--4089, 2009.

\bibitem{fridrikh1999polydomain}
S.V. Fridrikh and E.M. Terentjev.
\newblock Polydomain-monodomain transition in nematic elastomers.
\newblock {\em Physical Review E}, 60(2):1847, 1999.

\bibitem{verwey1997compositional}
G.C. Verwey and M.~Warner.
\newblock Compositional fluctuations and semisoftness in nematic elastomers.
\newblock {\em Macromolecules}, 30(14):4189--4195, 1997.

\bibitem{verwey1997nematic}
G.C. Verwey and M.~Warner.
\newblock Nematic elastomers cross-linked by rigid rod linkers.
\newblock {\em Macromolecules}, 30(14):4196--4204, 1997.

\bibitem{kutter2001tube}
S.~Kutter and E.M. Terentjev.
\newblock Tube model for the elasticity of entangled nematic rubbers.
\newblock {\em The European Physical Journal E}, 6:221--229, 2001.

\bibitem{plucinsky2017microstructure}
P.~Plucinsky and K.~Bhattacharya.
\newblock Microstructure-enabled control of wrinkling in nematic elastomer sheets.
\newblock {\em Journal of the Mechanics and Physics of Solids}, 102:125--150, 2017.

\bibitem{jeon2022synergistic}
S.Y. Jeon, B.~Shen, N.A. Traugutt, Z.~Zhu, L.~Fang, C.M. Yakacki, T.D. Nguyen, and S.H. Kang.
\newblock Synergistic energy absorption mechanisms of architected liquid crystal elastomers.
\newblock {\em Advanced Materials}, 34(14):2200272, 2022.

\bibitem{MAGHSOODI2023102060}
A.~Maghsoodi, M.O. Saed, E.M. Terentjev, and K.~Bhattacharya.
\newblock Softening of the hertz indentation contact in nematic elastomers.
\newblock {\em Extreme Mechanics Letters}, 63:102060, 2023.

\bibitem{farre2022dynamic}
H.J. Farre-Kaga, M.O. Saed, and E.M. Terentjev.
\newblock Dynamic pressure sensitive adhesion in nematic phase of liquid crystal elastomers.
\newblock {\em Advanced Functional Materials}, 32(12):2110190, 2022.

\bibitem{ohzono2019enhanced}
T.~Ohzono, M.O. Saed, and E.M. Terentjev.
\newblock Enhanced dynamic adhesion in nematic liquid crystal elastomers.
\newblock {\em Advanced Materials}, 31(30):1902642, 2019.

\bibitem{khaderi2015detachment}
S.N. Khaderi, N.A. Fleck, E.~Arzt, and R.M. McMeeking.
\newblock Detachment of an adhered micropillar from a dissimilar substrate.
\newblock {\em Journal of the Mechanics and Physics of Solids}, 75:159--183, 2015.

\bibitem{abaqus}
ABAQUS/Standard v.~6.20.
\newblock Dassault Systems Simulia Corp., Johnston, RI, USA, 2020.

\bibitem{LEE2023105369}
L.~Victoria, A.~Wihardja, and K.~Bhattacharya.
\newblock A macroscopic constitutive relation for isotropic-genesis, polydomain liquid crystal elastomers.
\newblock {\em Journal of the Mechanics and Physics of Solids}, 179:105369, 2023.

\bibitem{WTbook}
M.~Warner and E.M. Terentjev.
\newblock {\em Liquid Crystal Elastomers}.
\newblock Oxford University Press, 2007.

\bibitem{tanne}
E.~Tann\'e, T.~Li, B.~Bourdin, J.~J. Marigo, and C.~Maurini.
\newblock Crack nucleation in variational phase-field models of brittle fracture.
\newblock {\em Journal of the Mechanics and Physics of Solids}, 110:80--99, 2018.

\bibitem{hsueh}
C.J. Hsueh, L.~Avellar, B.~Bourdin, G.~Ravichandran, and K.~Bhattacharya.
\newblock Stress fluctuation, crack renucleation and toughening in layered materials.
\newblock {\em Journal of the Mechanics and Physics of Solids}, 120:68--78, 2018.

\bibitem{balijepalli2016numerical}
R.G. Balijepalli, M.R. Begley, N.A. Fleck, R.M. McMeeking, and E.~Arzt.
\newblock Numerical simulation of the edge stress singularity and the adhesion strength for compliant mushroom fibrils adhered to rigid substrates.
\newblock {\em International Journal of Solids and Structures}, 85:160--171, 2016.

\bibitem{balijepalli2017numerical}
R.G. Balijepalli, S.C. Fischer, R.~Hensel, R.M. McMeeking, and E.~Arzt.
\newblock Numerical study of adhesion enhancement by composite fibrils with soft tip layers.
\newblock {\em Journal of the Mechanics and Physics of Solids}, 99:357--378, 2017.

\bibitem{luo2020adhesion}
A.~Luo, A.~Mohammadi~Nasab, M.~Tatari, S.~Chen, W.~Shan, and K.T. Turner.
\newblock Adhesion of flat-ended pillars with non-circular contacts.
\newblock {\em Soft Matter}, 16(41):9534--9542, 2020.

\bibitem{gent}
A.N. Gent and P.B. Lindley.
\newblock Internal rupture of bonded rubber cylinders in tension.
\newblock {\em Proceedings of the Royal Society of London. Series A. Mathematical and Physical Sciences}, 249:195--205, 1959.

\bibitem{yao2006mechanics}
H.~Yao and H.~Gao.
\newblock Mechanics of robust and releasable adhesion in biology: Bottom--up designed hierarchical structures of gecko.
\newblock {\em Journal of the Mechanics and Physics of Solids}, 54(6):1120--1146, 2006.

\end{thebibliography}

 \end{document}